\documentclass[A4, graybox]{svmult}

\usepackage{stfloats}
\usepackage[normalem]{ulem}
\renewcommand{\uwave}[1]{#1}
\renewcommand{\sout}[1]{}

\usepackage{url}
\usepackage{mathptmx}       
\usepackage{helvet}         
\usepackage{courier}        
\usepackage{type1cm}        
\usepackage{makeidx}         
\usepackage{graphicx}        
\usepackage{multicol}        
\usepackage[bottom]{footmisc}

\usepackage{amsmath}
\usepackage{subfigure}
\usepackage{mathtools}

\makeindex             

\begin{document}

\title*{Communities unfolding in multislice networks}

\author{V. Carchiolo \and A. Longheu \and M. Malgeri \and G. Mangioni}
\institute{
Vincenza Carchiolo. Dipartimento di Ingegneria Informatica e delle
Telecomunicazioni, University of Catania, ITALY. E-mail: car@diit.unict.it
\and
Alessandro Longheu. Dipartimento di Ingegneria Informatica e delle
Telecomunicazioni, University of Catania, ITALY. E-mail: alongheu@diit.unict.it
\and
Michele Malgeri. Dipartimento di Ingegneria Informatica e delle
Telecomunicazioni, University of Catania, ITALY. E-mail: mm@diit.unict.it
\and Giuseppe Mangioni. Dipartimento di Ingegneria Informatica e delle
Telecomunicazioni, University of Catania, ITALY. E-mail: gmangioni@diit.unict.it
}

\maketitle

\abstract{Discovering communities in complex networks helps to
  understand the behaviour of the network. Some works in this
  promising research area exist, but communities uncovering in
  time-dependent and/or multiplex networks has not deeply investigated
  yet. In this paper, we propose a communities detection approach for
  multislice networks based on modularity optimization. We first
  present a method to reduce the network size that still preserves
  modularity. Then we introduce an algorithm that approximates
  modularity optimization (as usually adopted) for multislice
  networks, thus finding communities. The network size reduction
  allows us to maintain acceptable performances without affecting the
  effectiveness of the proposed approach.}

\section{Introduction}\label{s:introduction}
Communities structure detection in complex networks is a research
field that gained a considerable attention in the last few years. Such
interest is due to the possibility to discover hidden behaviours by
simply studying the network partitioning into communities. Several
methods to address the problem of community uncovering
(see~\cite{fortunato2010} for an overview) can be found in literature.
However, few of them consider the more general case of communities in
time-dependent networks
(\cite{chakrabarti2006}\cite{hopcroft2004}\cite{palla07}\cite{asur2007}\cite{fenn2009})
and/or multiplex networks. On the other hand, networks whose topology
evolves over time are quite common
(\cite{kumar2006}\cite{leskovec2009}). In this case, studying the
community structure by simply considering the network obtained by
adding together all of its snapshots over time can be too simplistic,
and it would not permit to investigate about the temporal evolution of
communities. To address this problem, recently Mucha et al.
\cite{mucha2010} presented a general framework to study the community
structure of arbitrary multislice networks, i.e. a set of individual
networks linked together by the use of inter-slice links.  Multislice
networks are general enough to be used to model time-varying,
multiplex and multiscale networks. To assess the quality of a given
partition into communities, Mucha et al. \cite{mucha2010} extended the
modularity function, originally introduced in
\cite{newman-2004-69}($Q_{NG}$), to be applied to the more general
case of multislice networks ($Q_{multislice}$).

A natural way to explore communities structure in multislice networks
is by direct optimization of the $Q_{multislice}$ function.
Unfortunately, \sout{it has been shown that} \uwave{the} exact
optimization of the $Q_{NG}$ modularity function is an $NP$-complete
problem~\cite{brandes08}, and \sout{a similar problem is also present
  in} the optimization of the $Q_{multislice}$ function
\uwave{presents a similar problem}. To deal with this problem, several
approximation methods have been developed (see \cite{fortunato2010}
for an overview). Among them, the Louvain method devised by Blondel
et al. \cite{blondel2008} is one of the fastest yet sufficiently
accurate algorithm.

In this work we present an algorithm inspired by \cite{blondel2008} to
discover communities in large multislice networks.  The paper is
organized as follows. In section~\ref{s:multislice} we introduce
multislice networks discussing about previous works in this
topic. Section~\ref{s:size_reduction} presents a method to reduce the
size of a multislice network while preserving modularity. Section
\ref{s:algorithm} illustrates our algorithm to discover community
structure in multislice networks. Finally, in
section~\ref{s:conclusions} conclusions and future works are
discussed.

\section{Communities in multislice networks}\label{s:multislice}
Real networks often are inherently dynamic, i.e. they change over
time.  Community structure in such networks cannot be effectively
analyzed neither only considering a single time snapshot nor studying
a new network obtained by a sort of ``sum'' of all the variations
across time. On the other hand, traditional approaches to community
discovering are not generally well suitable to manage multiplex (or
multi--layer) networks, where multiple edges between couple of nodes
are allowed. \emph{Multiplex} networks model different kind of
relations between nodes and can be, alternatively, represented as a
superimposition of distinct layers, each of which being the network
obtained by considering a single relation.

\begin{figure}
	\begin{center}
	\includegraphics[width=10cm,keepaspectratio=true]{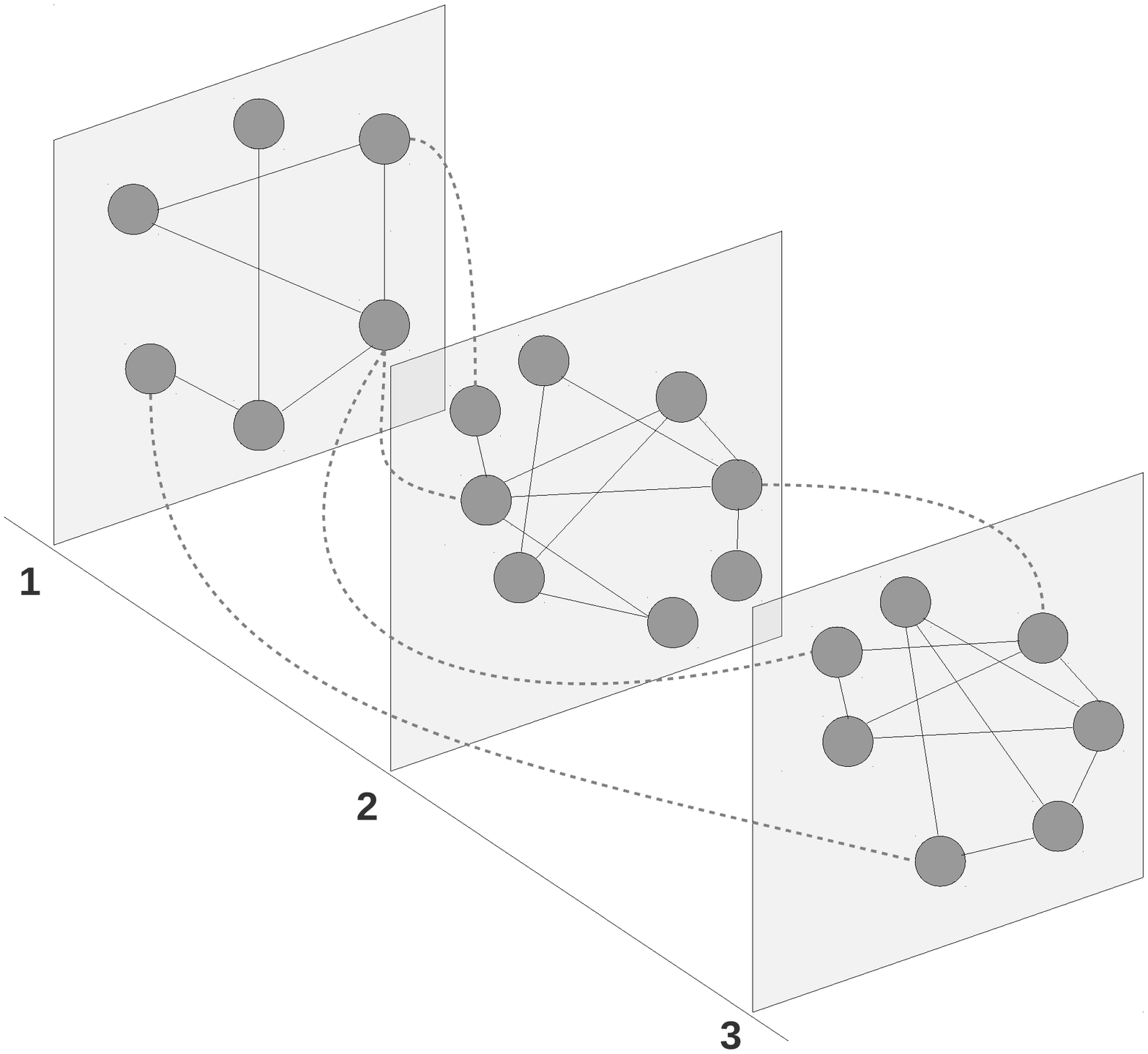}
	\caption{An example of a three slices network}
	\label{fig:multislice}
	\end{center}
\end{figure}

To address these issues, in \cite{mucha2010} the authors proposed a
framework to study community structure in multislice networks. A
\emph{multislice} is a network composed by a set of network slices
linked together by inter--slices links. As an example of such a
network, in figure~\ref{fig:multislice} it is reported a network
composed by three slices coupled each other by a set of links depicted
using dotted lines.
Multislice networks can be used in many contexts.  For instance, a
multiplex network can be simply represented by a multislice network by
mapping each layer of the network to a slice. Moreover, a time varying
network can be mapped to a multislice network where each slice is a
single instant snapshot network.

In \cite{mucha2010} the authors also propose a multislice extension of
the Newman's modularity function, thus providing a metric to assess
the quality of a given partition into communities of a multislice
network.

In particular, given a multislice network, the multislice
generalization of modularity for unipartite, undirected network slices
and couplings is:
\begin{equation}
  Q_{multislice}=\frac{1}{2\mu}\sum_{ijsr}\left\{ \left( A_{ijs} -\gamma_s\frac{k_{is}k_{js}}{2m_s}\right) \delta_{sr} + \delta_{ij}C_{jsr}
  \right\} \delta \left(g_{is},g_{jr}\right)
\label{eq:modularity}
\end{equation}
Where $i$ and $j$ range over all nodes, $s$ and $r$ range over all slices, 
$A_{ijs}$ is the element of the adjacency matrix of the slice
$s$ (intra--slice), $C_{jsr}$ is the link between node $j$ in slice
$s$ and node $j$ in slice $r$ (inter--slice coupling), $k_{is}$
($k_{js}$) is the degree of node $i$ ($j$) in slice $s$, $m_s$ is the
number of links in slice $s$, $\gamma_s$ is a resolution parameter and
$\mu$ is a normalization factor.

Equation~\ref{eq:modularity} can be considered as composed by two
terms, the first one takes into account the contribution to the
modularity given by each slice (it looks like Newman's modularity),
whereas the second term is the contribution given by the inter-slices
coupling.

The modularity function in equation~\ref{eq:modularity}
plays a double role: 1) it is used to
assess the quality of a given partition and 2) it can be exploited to
discover community structure by direct optimization.  Unfortunately,
as discussed in section \ref{s:introduction}, the exact modularity
optimization is presumably an $NP$-complete problem (similarly to what
has already been observed for Newman's modularity in
\cite{brandes08}). To overcome this computability matter, in this
paper we propose a greedy method to optimize $Q_{multislice}$ inspired
by the Louvain algorithm \cite{blondel2008}. In particular, our
algorithm makes extensively use of a network size reduction method
that we have specifically devised for multislice network, explained in
the next section.

\section{Size reduction in multislice networks}\label{s:size_reduction}
Reducing the size of multislice networks is useful to implement greedy
optimization for $Q_{multislice}$. To achieve this, let $G_m$ a
multislice network with undirected network slices and coupling (this
does not affect generality).

Note that, by definition, each node in slice $s$ is connected only
with the same node in slice $r$, that is $C_{ijsr}=0\;\forall i \neq
j$, then $C_{jsr} \equiv C_{ijsr}\;\forall i, j$. This equivalence
implies that the term $\delta_{ij}C_{jsr}$ in
equation\ref{eq:modularity} can be substituted by the equivalent
$C_{ijsr}$, so resulting equation is as follows:

\begin{equation}
  Q^{*}_{multislice}=\frac{1}{2\mu}\sum_{ijsr}\left\{ \left( A_{ijs}
      -\gamma_s\frac{k_{is}k_{js}}{2m_s}\right) \delta_{sr} + C_{ijsr}
  \right\} \delta \left(g_{is},g_{jr}\right)
\label{eq:qstar}
\end{equation}
Where $\delta_{ij}$ has been included into the coupling term.

By definition, for every partition into communities of $G_m
\Rightarrow Q^{*}_{multislice}\equiv Q_{multislice}$.

Now let $Com_s:\{1,...,N\}\rightarrow\{1,..M_s\}$ be a partition of
slice $s$ of the network into $M_s$ communities.  The function $Com_s$
assigns a community index $Com_s(i)$ to node $i$ in slice $s$ of the
network $G_m$.  Let us consider the reduced network $G'_m$ obtained as
in the following:
\begin{itemize}
\item In every slice $s$ we replaced each community with a single node.
\item The intra-slice weight $w'_{mns}$ between the nodes $m$ and $n$
  of slice $s$ of the reduced network $G'_m$ is defined as in the
  following:
  \begin{equation}
    w'_{mns} = \sum_i\sum_j A_{ijs}\delta(Com_s(i), m)\delta(Com_s(j),n) 
    \; m,n \in {1,...,M_s}
    \label{eq:reduced_intraslices}
  \end{equation}
  i.e. $w'_{mns}$ is the sum of all the links connecting vertices in
  the corresponding communities.
\item The inter-slice weight $C'_{mnsr}$ between node $m$ in slice $s$
  and node $n$ in slice $r$ of the reduced network $G'_m$ is defined
  as in the following:
  \begin{equation}
    C'_{mnsr} = \sum_i\sum_j C_{ijsr}\delta(Com_s(i), m)\delta(Com_r(j),n) 
    \; m \in {1,...,M_s},\; n \in {1,...,M_r}
    \label{eq:reduced_interslices}
  \end{equation}
  i.e. $C'_{mnsr}$ is the sum of all the links connecting vertices in
  community $m$ in slice $s$ with vertices in community $n$ in slice
  $r$.
\end{itemize}	
In other words, the reduced multislice network is obtained by
collapsing each community in one node and by properly setting the
weights of both inter--slice and intra--slice links.

\begin{figure}
  \centering \subfigure[Original multislice network - $G_m$]
  {\includegraphics[width=6.4cm,keepaspectratio=true]{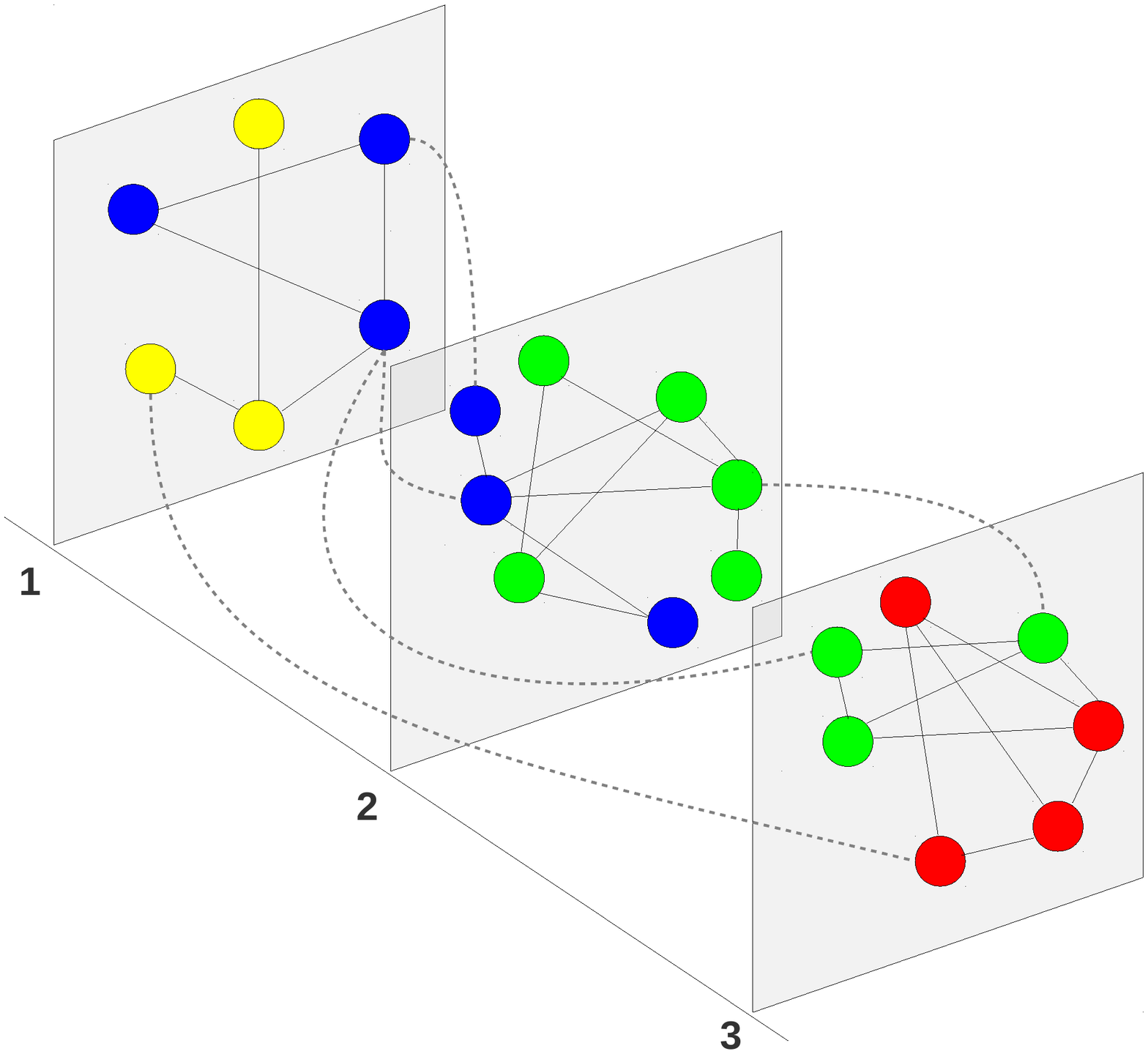}\label{fig:reduction1}
  } \hspace{-1.5cm} \subfigure[Reduced multislice network - $G^{'}_m$]
  {\includegraphics[width=6.4cm,keepaspectratio=true]{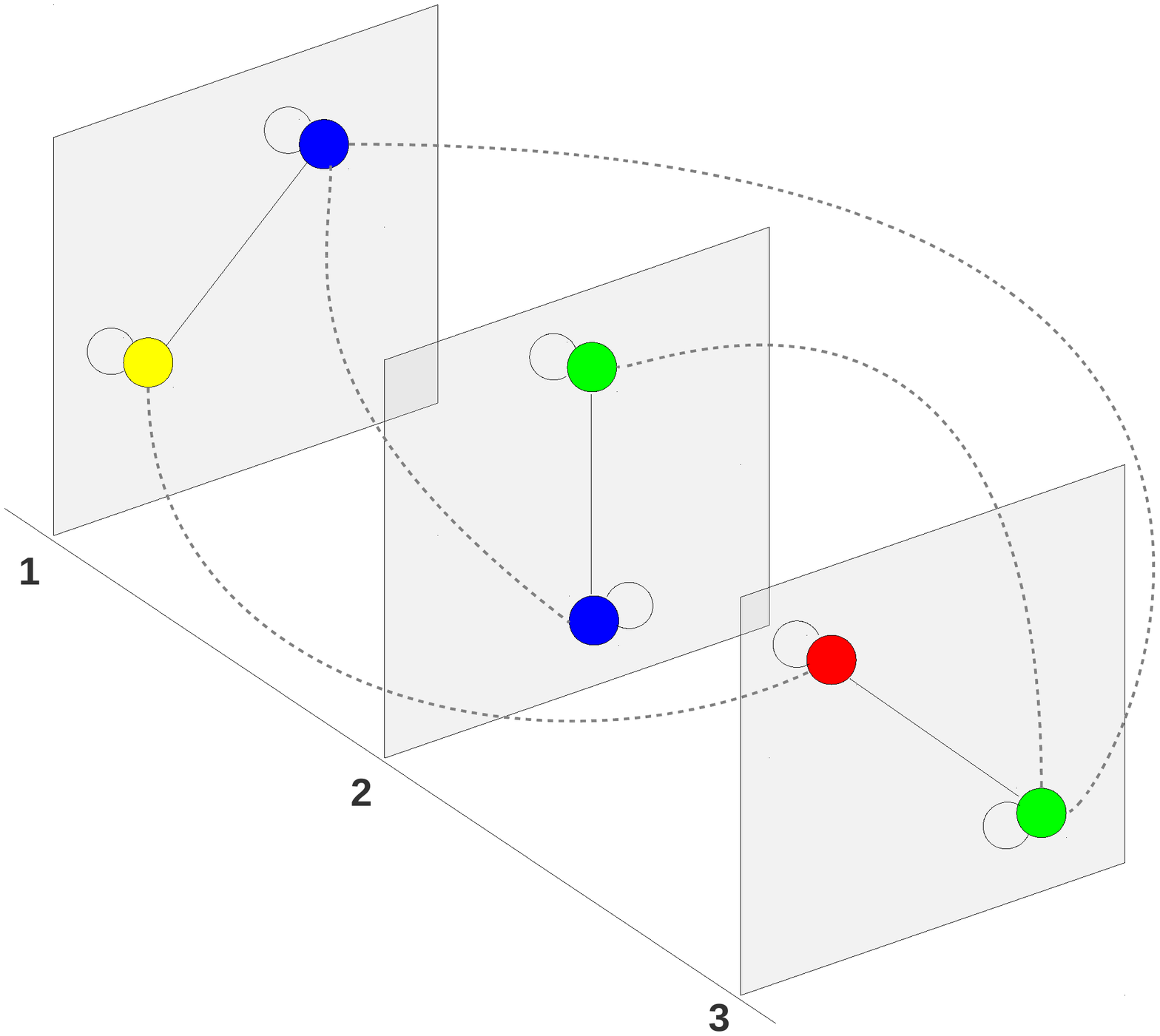}\label{fig:reduction2}
  }
  \label{fig:reduction}
\end{figure}

\sout{In} Figure~\ref{fig:reduction} \uwave{presents} an example of the
application of the proposed size reduction method \sout{is
shown}. Figure~\ref{fig:reduction1} shows the original multislice
network $G_m$ composed by three slices where nodes belonging to a
community are depicted using the same colour.
Figure~\ref{fig:reduction2} shows the reduced multislice network
$G^{'}_{m}$ composed (as the original one) of three slices, where each
community has been replaced by one node and link weights are
recomputed by using equations~\ref{eq:reduced_intraslices} and
\ref{eq:reduced_interslices}.

Now we want to \sout{show}\uwave{prove} that the $Q^{*}_{multislice}$
of $G_m$ is equal to $Q^{*'}_{multislice}$ of $G'_m$, i.e. the
proposed network size reduction method preserves multislice
modularity.

The proof that $Q^{*'}_{multislice} = Q^{*}_{multislice}$ is as follow:
\begin{align}
	\label{eq:q_equiv1}
  Q^{*'}_{multislice}=&\frac{1}{2\mu'}\sum_{mnsr}\left\{ \left(
      w'_{mns} -\gamma_s\frac{w'_{ms}w'_{ns}}{2w'_s}\right)
    \delta_{sr} + C'_{mnsr}
  \right\} \delta \left(g_{ms},g_{nr}\right) \notag \\  
  =&\underbracket{\frac{1}{2\mu'}\sum_{s}\sum_{mn} \left( w'_{mns}
      -\gamma_s\frac{w'_{ms}w'_{ns}}{2w'_s}\right) \delta
    \left(g_{ms},g_{ns}\right)}_{\text{1st term}} + \\
  &+ \underbracket{\frac{1}{2\mu'}\sum_{sr}\sum_{mn}C'_{mnsr} \delta
    \left(g_{ms},g_{nr}\right)}_{\text{2nd term}} \notag
\end{align}

By applying the same approach followed in \cite{arenas-2007-9}, it is
straightforward to prove that the first term in
equation~\ref{eq:q_equiv1} can be rewritten as:
\begin{equation}
  1st\;term=\frac{1}{2\mu}\sum_{s}\sum_{ij} \left( w_{ijs}
    -\gamma_s\frac{w_{is}w_{js}}{2w_s}\right) \delta
  \left(g_{is},g_{js}\right)
\label{eq:1term}
\end{equation}

By using equation~\ref{eq:reduced_interslices}, it is also easy to
show that the second term can be rewritten as:
\begin{align}
  2nd\;term=&\frac{1}{2\mu'}\sum_{sr}\sum_{mn}C'_{mnsr} \delta
  \left(g_{ms},g_{nr}\right) \notag \\
  =& \frac{1}{2\mu}\sum_{sr}\sum_{mn}\left(\sum_{ij}
    C_{ijsr}\delta(Com_s(i), m)\delta(Com_r(j),n)\right) \delta
  \left(g_{ms},g_{nr}\right) \notag \\
  =& \frac{1}{2\mu}\sum_{sr}\sum_{ij}C_{ijsr}\sum_{mn}
  \delta(Com_s(i), m)\delta(Com_r(j),n) \delta
  \left(g_{ms},g_{nr}\right)\\
  =& \frac{1}{2\mu}\sum_{sr}\sum_{ij}C_{ijsr}
  \delta\left(g_{Com_s(i)s},g_{Com_r(j)r}\right) \notag \\
  =& \frac{1}{2\mu}\sum_{sr}\sum_{ij}C_{ijsr} \delta\left(g_{is},g_{jr}\right) \notag
\label{eq:2term}
\end{align}

Putting together the first and second terms, we obtain the following:
\begin{align}
  Q^{*'}_{multislice}=&\underbracket{\frac{1}{2\mu}\sum_{s}\sum_{ij}
    \left( w_{ijs}-\gamma_s\frac{w_{is}w_{js}}{2w_s}\right) \delta
    \left(g_{is},g_{js}\right)}_{\text{1st term}}   
 + \underbracket{\frac{1}{2\mu}\sum_{sr}\sum_{ij}C_{ijsr} \delta
\left(g_{is},g_{jr}\right)}_{\text{2nd term}} \notag \\
			=&\frac{1}{2\mu}\sum_{ijsr}\left\{ \left( w_{ijs}
-\gamma_s\frac{w_{is}w_{js}}{2w_s}\right) \delta_{sr} + C_{ijsr}
\right\} \delta \left(g_{is},g_{jr}\right)  \\
			=& Q^{*}_{multislice} \notag
\label{eq:q_12term}
\end{align}

In conclusion we
proved that nodes belonging to a community in a multislice network can
be all replaced by a unique node in the reduced multislice network (
this is a generalization of the work by Arenas et al. in
\cite{arenas-2007-9}).
	
\section{An algorithm to discover communities in multislice
networks}\label{s:algorithm}
To discover communities in multislice networks we propose a greedy
algorithm based on a local optimization of the modularity function in
equation~\ref{eq:qstar}.  

Given a multislice network $G_m$ , our algorithm consists of two steps
repeated iteratively:

\subsubsection*{Step 1 }

\begin{itemize}
\item Initially, we place each node of the network in a different
  community, so there are as many communities as the nodes in the
  multislice network (i.e. $\sum_s N_s$ where $N_s$ is the number of
  nodes in slice $s$).
\item For each node $i$ in \sout{each} \uwave{the} slice $s$ the gain of
  $Q^{*}_{multislice}$ obtained by moving node $i$ in the same
  community of it's neighbours $j$ is computed. Note that the
  neighbourhood of a node $i$ is composed by all nodes $i$ is linked
  to. It also includes those nodes $i$ is linked to through inter-slices
  coupling.
\item Then, node $i$ is placed in the community for which the gain is
  maximum (and positive).
\item Step 1 is performed iteratively until a local maximum of
  $Q^{*}_{multislice}$ is reached. 
\end{itemize}

\subsubsection*{Step 2}

  After step 1, we build a new multislice network by applying the size
  reduction method described in section~\ref{s:size_reduction}.
  \begin{itemize}
  \item Each slice of the new network consists of as many nodes as the
    number of communities found during the step 1.
  \item The weight of the intra-slice link between two new nodes $i$
    and $j$ is given by the sum of the weights of the links between
    communities corresponding to nodes $i$ and $j$ respectively
    (eq.~\ref{eq:reduced_intraslices}).  Note that intra-slice links
    between nodes in the same community are represented by a weighted
    self--loop link in the corresponding new node.
  \item The weight $C_{ijsr}$ of the inter-slice links between node
    $i$ in slice $s$ and node $j$ in slice $r$ is given by the sum of
    the weights of the links between communities corresponding to
    nodes $i$ and $j$ placed in slices $s$ and $r$ respectively
    (eq.~\ref{eq:reduced_interslices}).
\end{itemize}

\begin{figure}
  \begin{center}
    \includegraphics[width=14cm,keepaspectratio=true]{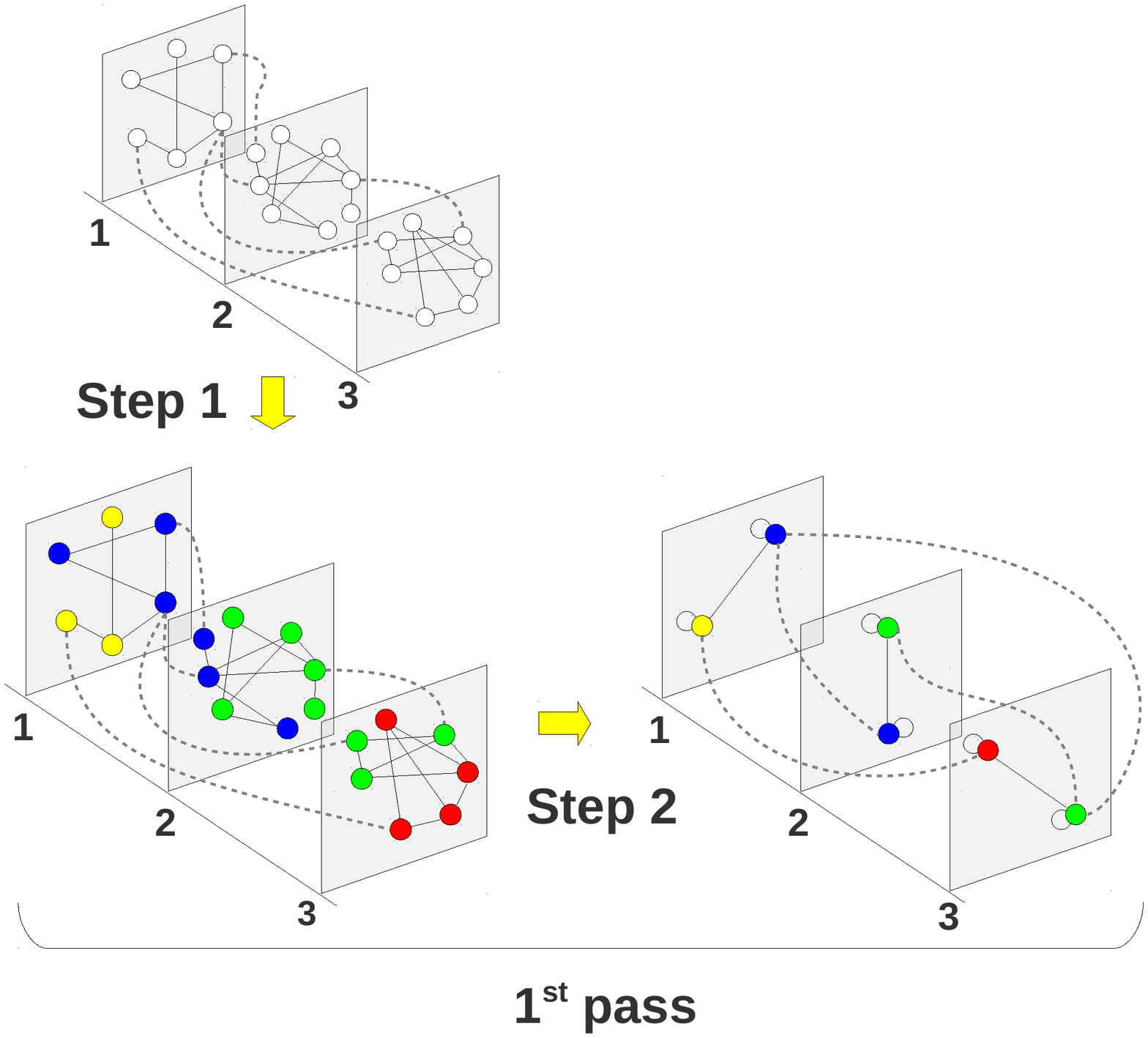}
  \end{center}
  \label{fig:alg}
  \caption{}
\end{figure}

After the second step the number of nodes can diminish drastically,
thus speeding up the computation time.  To get an idea of how much
network size decreases thanks to the proposed reduction method,
readers can refer to the work by Arenas et al.~\cite{arenas-2007-9}.
In figure~\ref{fig:alg} the two steps of our algorithm are graphically
illustrated for a network composed by three slices.

In addition, the way the algorithm works permits an implicit
discovering of the hierarchical structure of a multislice network. In
fact, the network produced at the end of the second step in each pass
of the algorithm can be considered as a more higher hierarchical level
network.  In other words, the hierarchical organization of the network
is naturally explored as the algorithm proceeds.  In conclusion, our
algorithm inherits all the advantages of the Louvain method proposed
by V.  Blondel et al.\cite{blondel2008}:
\begin{itemize}
\item It is easy to implement.
\item It is very fast (Blondel et al.\cite{blondel2008}
  claim that their implementation is able to find communities in a network
  of 118 million nodes and 1 billion links in 152 minutes only!)
\item It is multi-resolution and naturally gives a hierarchical
  decomposition of the network.
\end{itemize}
Additionally, our proposal
also allows us to discover communities in multislice networks.

To test our algorithm, we have implemented a prototype written in
Python programming language.  Since no universally accepted benchmarks
for multislice networks currently exist we performed our tests on the
examples provided in \cite{mucha2010} and specifically we discovered
communities in the Zachary Karate network across multiple
resolutions. Running our algorithm on this network we obtained the
same results reported in \cite{mucha2010}, thus endorsing the
effectiveness of the proposed method.

\section{Conclusions}\label{s:conclusions}
In this paper we presented a greedy algorithm to find communities in
multislice networks. Our proposal started from reducing the size of
the network without affecting the modularity, so the reduced network
partitioning into communities is equivalent to the initial network.

Some issues still have to be addressed, in particular:
\begin{itemize}
\item to replace the prototype in Python with an optimized C++
  implementation of the greedy algorithm described previously
\item to test the proposed approach to real and large networks, in
  order to assess its effectiveness as well as its performances. This
  goal is strictly related to the previous one, indeed an optimized
  algorithm implementation is essential when working on large networks
\item to investigate about the definition of (new) benchmarks for
  multislice networks in addition to currently available benchmarks as
  \cite{lancichinetti2009}\cite{lancichinetti2008}\cite{girvan2002}.
\end{itemize}

\bibliography{biblio}

\begin{thebibliography}{10}

\bibitem{arenas-2007-9}
A.~Arenas, J.~Duch, A.~Fernandez, and S.~Gomez.
\newblock Size reduction of complex networks preserving modularity.
\newblock {\em New Journal of Physics}, 9:176, 2007.

\bibitem{asur2007}
Sitaram Asur, Srinivasan Parthasarathy, and Duygu Ucar.
\newblock An event-based framework for characterizing the evolutionary behavior
  of interaction graphs.
\newblock In {\em Proceedings of the 13th ACM SIGKDD international conference
  on Knowledge discovery and data mining}, KDD '07, pages 913--921, New York,
  NY, USA, 2007. ACM.

\bibitem{blondel2008}
Vincent~D Blondel, Jean-Loup Guillaume, Renaud Lambiotte, and Etienne Lefebvre.
\newblock Fast unfolding of communities in large networks.
\newblock {\em Journal of Statistical Mechanics: Theory and Experiment},
  2008(10):P10008, 2008.

\bibitem{brandes08}
U.~Brandes, D.~Delling, M.~Gaertler, R.~Gorke, M.~Hoefer, Z.~Nikoloski, and
  D.~Wagner.
\newblock On modularity clustering.
\newblock {\em Knowledge and Data Engineering, IEEE Transactions on},
  20(2):172--188, Feb. 2008.

\bibitem{chakrabarti2006}
Deepayan Chakrabarti, Ravi Kumar, and Andrew Tomkins.
\newblock Evolutionary clustering.
\newblock In {\em Proceedings of the 12th ACM SIGKDD international conference
  on Knowledge discovery and data mining}, KDD '06, pages 554--560, New York,
  NY, USA, 2006. ACM.

\bibitem{fenn2009}
D.~J. {Fenn}, M.~A. {Porter}, M.~{McDonald}, S.~{Williams}, N.~F. {Johnson},
  and N.~S. {Jones}.
\newblock {Dynamic communities in multichannel data: An application to the
  foreign exchange market during the 2007-2008 credit crisis}.
\newblock {\em Chaos}, 19(3):033119--+, September 2009.

\bibitem{fortunato2010}
Santo Fortunato.
\newblock Community detection in graphs.
\newblock {\em Physics Reports}, 486:75--174, 2010.

\bibitem{girvan2002}
M.~Girvan and M.~E.~J. Newman.
\newblock {Community structure in social and biological networks}.
\newblock {\em Proceedings of the National Academy of Sciences of the United
  States of America}, 99(12):7821--7826, June 2002.

\bibitem{hopcroft2004}
John Hopcroft, Omar Khan, Brian Kulis, and Bart Selman.
\newblock {Tracking evolving communities in large linked networks}.
\newblock {\em Proceedings of the National Academy of Sciences},
  101:5249--5253, April.

\bibitem{kumar2006}
Ravi Kumar, Jasmine Novak, and Andrew Tomkins.
\newblock Structure and evolution of online social networks.
\newblock In {\em Proceedings of the 12th ACM SIGKDD international conference
  on Knowledge discovery and data mining}, KDD '06, pages 611--617, New York,
  NY, USA, 2006. ACM.

\bibitem{lancichinetti2009}
Andrea Lancichinetti and Santo Fortunato.
\newblock Benchmarks for testing community detection algorithms on directed and
  weighted graphs with overlapping communities.
\newblock {\em Phys. Rev. E}, 80(1):016118, Jul 2009.

\bibitem{lancichinetti2008}
Andrea Lancichinetti, Santo Fortunato, and Filippo Radicchi.
\newblock Benchmark graphs for testing community detection algorithms.
\newblock {\em Phys. Rev. E}, 78(4):046110, October 2008.

\bibitem{leskovec2009}
Jure Leskovec, Kevin~J. Lang, Anirban Dasgupta, and Michael~W. Mahoney.
\newblock Community structure in large networks: Natural cluster sizes and the
  absence of large well-defined clusters.
\newblock {\em Internet Mathematics}, 6(1):29--123, 2009.

\bibitem{mucha2010}
Peter~J. Mucha, Thomas Richardson, Kevin Macon, Mason~A. Porter, and
  Jukka-Pekka Onnela.
\newblock Community structure in time-dependent, multiscale, and multiplex
  networks.
\newblock {\em Science}, 328(5980):876--878, May 2010.

\bibitem{newman-2004-69}
M.~E.~J. Newman and M.~Girvan.
\newblock Finding and evaluating community structure in networks.
\newblock {\em Physical Review E}, 69:026113, 2004.

\bibitem{palla07}
Gergely Palla, Albert lászló Barabási, Tamás Vicsek, and Budapest Hungary.
\newblock Quantifying social group evolution.
\newblock {\em Nature}, 446:2007, 2007.

\end{thebibliography}
\bibliographystyle{plain}

\end{document}